\documentclass[11pt,superscriptaddress,aps,prd,preprint]{revtex4}
\usepackage{amsmath}

\makeatletter
\usepackage[T1]{fontenc}
\usepackage{amsmath}
\usepackage{amssymb}
\usepackage{graphicx}

\newcommand{\bea}{\begin{eqnarray}}
\newcommand{\eea}{\end{eqnarray}}

\begin{document}
\title{G\"{o}del-type solution in $f(R,T)$ modified gravity}

\author{C. J. Ferst}\email[]{celso.ferst@fisica.ufmt.br}
\affiliation{Instituto de F\'{\i}sica, Universidade Federal de Mato Grosso,\\
78060-900, Cuiab\'{a}, Mato Grosso, Brazil}

\author{A. F. Santos}\email[]{alesandroferreira@fisica.ufmt.br}
\affiliation{Instituto de F\'{\i}sica, Universidade Federal de Mato Grosso,\\
78060-900, Cuiab\'{a}, Mato Grosso, Brazil}

\begin{abstract}
In this paper, we will examine the problem of violation of causality in $f(R,T)$ modified gravity, where $R$ is the Ricci scalar and $T$ is the trace of the energy-momentum tensor $T_{\mu\nu}$. We investigate the causality problem in two special cases, in the first we consider the matter content of the universe as a perfect fluid and in the second case, the matter content is a perfect fluid plus a scalar field.
\end{abstract}

\maketitle
Observational data confirm that the universe is expanding at an accelerated rate \cite{Riess, Perm, Riess2}. In recent years, a lot of researchers are in search to answer the question: what causes this acceleration in the universe? Currently, there are two different possibilities that attempt to explain this fact: one indicates that the agent responsible for the accelerated expansion of the universe is a component of the energy density of the universe, which is approximately 68\% of the energy density of the universe \cite{Planck}, called dark energy, and the other possibility is to modify the gravitational theory, i.e., modify the general relativity. There are numerous models of gravitation alternative to general relativity, called modified theories of gravity, as an example we can mention the Chern-Simons gravity \cite{CS, CS1}, where the modification consists of adding a term that leads to parity violation, another model is the Einstein-Aether  gravitational model \cite{Ted, Ted1}, that is a model that displays Lorentz symmetry violation. We can as well cite the Horava-Lifshitz gravity \cite{Horava} an anisotropic gravitational theory, where is introduced the Lifshitz symmetry which promotes anisotropy between space and time.

A model that has attracted attention in recent years is $f(R)$ gravity, a revision of this model can be found in \cite{Felice, Clifton, Capo}. This theory arises when we replace in the Einstein-Hilbert action the Ricci scalar $R$ by any function $f(R)$. It offers an alternative means to explain the present cosmic acceleration without involving any exotic component, i.e., dark energy, or the existence of additional spatial dimensions. Another theory is $f(R,T)$ gravity, which is a generalization of $f(R)$ gravity and was first introduced in \cite{Harko}. In this case, the Ricci scalar $R$ in the Hilbert-Einstein action, is changed by some function that dependent of the Ricci scalar $R$ and of $T$, the trace of the energy-momentum tensor $T_{\mu\nu}$. The dependence of $T$ can be induced by imperfect fluids or conformal anomalies coming from quantum effects. This gravitational model has been intensively investigated in recent years, as examples we can mention: some cosmological aspects have been investigated in \cite{Houndjo, Momeni, Chatto, Shabani}, in \cite{Alvarenga, Sharif, Kiani} the energy conditions were considered, the thermodynamic properties were analyzed in \cite{Houndjo2, Sharif2, Jamil, Sharif3}, in \cite{Azizi} wormhole solution was examined in the setting of the $f(R,T)$ gravity, other interesting discussions can be seen in \cite{Reddy, Baffou, Moraes}. 

In the work \cite{Santos} was investigated the G\"{o}del universe in the $f(R,T)$ modified gravity theory, where it was shown that for a given choice of the cosmological constant and of the energy density this solution also takes place in the $f(R,T)$ gravity as well as in general relativity. The G\"{o}del universe, proposed in 1949 \cite{Godel}, is a model of universe that has as main feature the possibility of closed time-like curves (CTC's), i.e., at least in theory, it would be possible to have time travel, but specifically, turn back to the past. In this paper, we consider a G\"{o}del-type metric in the context of gravity $f(R,T)$ in order to generalize the study realized in \cite{Santos} and the study done in \cite{Reboucas1} for $f(R)$ gravity, which is a generalization of the study realized in \cite{Barrow}. We want to determine the critical radius, which determines the existence of CTC's, for two cases: (i) content of matter, it is only a perfect fluid and (ii) matter is composed of a perfect fluid plus a scalar field.

The action that describes this gravity model is given by \cite{Harko}
\bea
S=\frac{1}{16\pi G}\int \sqrt{-g}\left[f(R,T)+16\pi G{\cal L}_m\right]d^4x,
\eea
where $f(R,T)$ is a function of the Ricci scalar $R$ and of the trace of the energy-momentum tensor $T$. If $f(R,T)=R$, this for the case in which $T=0$, we recover the usual Einstein-Hilbert action. ${\cal L}_m$ is the Lagrangian of matter, from which we can define the energy momentum tensor as
\bea
T_{\mu\nu}=-\frac{2}{\sqrt{-g}}\frac{\delta(\sqrt{-g}{\cal L}_m)}{\delta g^{\mu\nu}}.
\eea
Assuming that ${\cal L}_m$ only depends on the components $g_{\mu\nu}$ and using the fact that
\bea
\delta\sqrt{-g}&=&\frac{1}{2}\sqrt{-g}g^{\mu\nu}\delta g_{\mu\nu},
\eea
and that
\bea
g^{\mu\nu}\delta g_{\mu\nu}=-g_{\mu\nu}\delta g^{\mu\nu}, 
\eea
we find for the energy-momentum tensor that
\bea
T_{\mu\nu}=g_{\mu\nu}{\cal L}_m-2\frac{\partial{\cal L}_m}{\partial g^{\mu\nu}}.\label{TEM}
\eea

Varying the action with respect to the metric tensor $g^{\mu\nu}$ we find the equations of motion as
\bea
&&f_R(R,T)R_{\mu\nu}-\frac{1}{2}f(R,T)g_{\mu\nu}+(g_{\mu\nu}\Box-\nabla_\mu\nabla_\nu)f_R(R,T)\nonumber\\
&&=8\pi G T_{\mu\nu}-f_T(R,T)T_{\mu\nu}-f_T(R,T)\Theta_{\mu\nu},\label{mov}
\eea
where $R_{\mu\nu}$ is the Ricci tensor, $f_R(R,T)=\frac{\partial f(R,T)}{\partial R}$, $f_T(R,T)=\frac{\partial f(R,T)}{\partial T}$ and 
\bea
\Theta_{\mu\nu}\equiv g^{\lambda\rho}\frac{\delta T_{\lambda\rho}}{\delta g^{\mu\nu}}.\label{Theta}
\eea
To obtain a relation between the Ricci scalar and the trace of the energy-momentum tensor we contract the equation (\ref{mov}) so that we find
\bea
f_R(R,T)R+3\Box f_R(R,T)-2f(R,T)=8\pi G T-f_T(R,T)T-f_T(R,T)\Theta,
\eea
where $\Theta=g^{\mu\nu}\Theta_{\mu\nu}$ is the trace of the tensor $\Theta_{\mu\nu}$.

Using the equation (\ref{TEM}) we can write the tensor $\Theta_{\mu\nu}$, equation (\ref{Theta}), as
\bea
\Theta_{\mu\nu}=-2T_{\mu\nu}+g_{\mu\nu}{\cal L}_m-2g^{\alpha\beta}\frac{\partial^2{\cal L}_m}{\partial g^{\mu\nu}\partial g^{\alpha\beta}}.
\eea
Assuming that the matter content is a perfect fluid, whose energy momentum tensor is given by
\bea
T_{\mu\nu}=(\rho+p)u_\mu u_\nu-pg_{\mu\nu},
\eea
and that the Lagrangian of matter can be given by
\bea
{\cal L}_m=-p,
\eea
we get
\bea
\Theta_{\mu\nu}=-2T_{\mu\nu}-pg_{\mu\nu}.\label{Theta1}
\eea

Now we will discuss the problem of causality in this gravity model, in the context of space-time of G\"{o}del, more specifically G\"{o}del-type metric given by \cite{Reboucas1, Reboucas2}
\bea
ds^2=\left[dt+H(r)d\phi\right]^2-D^2(r)d\phi^2-dr^2-dz^2,\label{godel-type}
\eea
where
\bea
H(r)&=&\frac{4\omega}{m^2}senh^2\left(\frac{mr}{2}\right),\label{H}\\
D(r)&=&\frac{1}{m}senh\left(mr\right)\label{D}.
\eea
The parameters $\omega$ and $m$ characterize all types of G\"{o}del-types metrics. We can rewrite the metric (\ref{godel-type}) as
\bea
ds^2=dt^2+2H(r)dtd\phi-dr^2-G(r)d\phi^2-dz^2,
\eea
with $G(r)=D^2(r)-H^2(r)$. Here it is observed that the existence of closed time-like curves is related to the behavior of the function $G(r)$ and from which one can determine a critical radius $r_c$, in an interval said $r_1<r<r_2$, beyond which closed time-like curves (CTC's) exist \cite{Reboucas1}. Using the functions (\ref{H}) and (\ref{D}) in the function $G(r)$ we find that \cite{Reboucas2}
\bea
senh^2\left(\frac{mr_c}{2}\right)=\left(\frac{4\omega^2}{m^2}-1\right)^{-1}.\label{senh}
\eea

We can analyze the equation (\ref{senh}) for two important cases, which relate the parameters of the geometry $\omega$ and $m$: 

{\it Case I:} $m^2=2\omega^2$

In this case we find
\bea
senh^2\left(\frac{mr_c}{2}\right)&=&\left(\frac{4\omega^2}{2\omega^2}-1\right)^{-1},\nonumber\\
r_c&=&\frac{2}{m}senh^{-1}(1),\label{raio}
\eea
a similar result to that obtained by self K. G\"{o}del in 1949 \cite{Godel}, i.e., we obtain a finite critical radius and thus we have a violation of causality, since we have a region of space-time that exhibits closed time-like curves.

{\it Case II:} $m^2\geq 4\omega^2$

In this case, the region where the causality is violated is avoided because when we substitute $m^2=4\omega^2$ in the equation (\ref{senh}) we obtain
\bea
r_c\to + \infty.
\eea

With the purpose of simplifying calculations with equations of motion we use the choice of base proposal in \cite{Reboucas2},
\bea
ds^2=\eta_{AB}\theta^A\theta^B,\label{base}
\eea
where $\eta_{AB}=diag(+,-,-,-)$ is a flat metric and
\bea
\theta^0&=&dt+H(r)d\phi,\,\,\,\,\,\,\,\,\,\,\,\,\,\,\,\,\,\,\,\, \theta^1=dr,\nonumber\\
\theta^2&=&D(r)d\phi,\,\,\,\,\,\,\,\,\,\,\,\,\,\,\,\,\,\,\,\,\,\,\,\,\,\,\,\,\,\,\theta^3=dz.
\eea

With this choice and noting that from the geometry (\ref{godel-type}) we find that the derivatives of the function $ f_R $ vanish because $R=2(m^2-\omega^2)$, we can rewrite the equations of motion as
\bea
f_R G_{AB}=\kappa^2T_{AB}-f_T(T_{AB}+\Theta_{AB})-\frac{1}{2}\left[f+\kappa^2 T-f_T(T+\Theta)\right]\eta_{AB}.\label{motion2}
\eea
Now let's analyze the equations of motion for a given content of matter and energy. First, let's consider that the source of matter is a perfect fluid whose energy-momentum tensor, in the base chosen in (\ref{base}), is given by
\bea
T_{AB}=(\rho+p)u_Au_B-p\eta_{AB},\label{tensorM}
\eea
where $\rho$ is the energy density, $p$ is the pressure and $u_A=(1,0,0,0)$ is the four-velocity of the fluid. From the expression of the energy-momentum tensor we can calculate the trace of this quantity given by
\bea
T=\eta^{AB}T_{AB}=\rho-3p.
\eea
Yet in this basis, we can rewrite the tensor (\ref{Theta1}) as
\bea
\Theta_{AB}=-2T_{AB}-p\eta_{AB},
\eea
and from it we find that the trace of this tensor is given by
\bea
\Theta=\Theta_{AB}\eta^{AB}=2(p-\rho).
\eea

Substituting these last relations in the equation of motion (\ref{motion2}) we stay with
\bea
f_RG_{AB}=\kappa^2 T_{AB}+f_T(\rho+p)u_Au_B-\frac{1}{2}\left[f+\kappa^2 T+f_T(\rho+p)\right]\eta_{AB}.
\eea
To write all components of the equation of motion  above, we find that the non-zero components of the Einstein tensor are given by
\bea
G_{00}&=&3\omega^2-m^2,\nonumber\\
G_{11}&=&G_{22}=\omega^2,\nonumber\\
G_{33}&=&m^2-\omega^2,\label{einsteintensor}
\eea
and thus the field equations that result are:
\bea
2f_R(3\omega^2-m^2)+f&=&\kappa^2(\rho+3p)+f_T(\rho+p),\label{eq1}\\
2f_R\omega^2 -f &=&\kappa^2(\rho-p)+f_T(\rho+p),\label{eq2}\\
2f_R(m^2-\omega^2)-f&=&\kappa^2(\rho-p)+f_T(\rho+p).\label{eq3}
\eea
Rearranging the equations (\ref{eq2}) and (\ref{eq3}) we obtain
\bea
(2\omega^2-m^2)f_R=0,\label{eq4}
\eea
an equation that is similar to that obtained in \cite{Reboucas1} for the $f(R)$ gravity. We note that, to have a  $f(R,T)$ theory, we assume that $f_R\neq 0$, and in this case we obtain that
\bea
2\omega^2=m^2,\label{violate}
\eea
Which leads us directly to the G\"{o}del metric, case I explained above, and so we can say that there is a violation of causality in this gravity model when choosing the perfect fluid as matter content of the universe. Once this situation displays violation of causality, we can determine the critical radius, $r_c$, which defines the  closed time-like curves, and investigate the dependence of this function with $f(R,T)$ and its derivatives with respect to $R$ and $T$. Manipulating the equations of fields (\ref{eq1}), (\ref{eq2}) and (\ref{eq3}) we stay with 
\bea
m^2f_R+f&=&\kappa^2(\rho+3p)+f_T(\rho+p),\label{eq5}\\
m^2f_R-f&=&\kappa^2(\rho-p)+f_T(\rho+p).\label{eq6}
\eea
And of these we find that 
\bea
\frac{1}{2}f=\kappa^2\,p.\label{eq7}
\eea
Lastly, if we replace the equation (\ref{eq7}) in the equation (\ref{eq5}), or in the equation (\ref{eq6}), we obtain
\bea
m=\sqrt{\frac{2\kappa^2\rho+f+2f_T(\rho+p)}{2f_R}}.
\eea
The critical radius, equation (\ref{raio}), in this case is given by
\bea
r_c=2\,senh^{-1}(1)\sqrt{\frac{2f_R}{2\kappa^2\rho+f+2f_T(\rho+p)}}.
\eea
With this result, we see that the critical radius, beyond which causality violation occurs explicitly depends on the form of the function $f(R,T)$ and its derivatives, $f_R$ and $f_T$, with respect to $R$ and $T$, respectively. With this, we note that this result is a generalization of the case obtained for $f(R) $ gravity.

Until here we saw that, when the content of matter is only a perfect fluid the violation of causality  arises naturally, of a form similar to that occurs in the $f(R)$ gravity. In the attempt of we find a causal solution for the G\"{o}del-type metric in the $f(R,T)$ gravity, we go consider that the distribution of matter is composed of two parts, i.e.,
\bea
T_{AB}=T_{AB}^M+T_{AB}^S,
\eea
where $T_{AB}^M$ is the energy-momentum tensor associated with the perfect fluid given by equation (\ref{tensorM}) and $T_{AB}^S$ is the energy-momentum tensor associated to a scalar field given by
\bea
T_{AB}^S=\nabla_A\Phi\nabla_B\Phi-\frac{1}{2}\eta_{AB}\,\eta^{CD}\nabla_C\Phi\nabla_D\Phi,\label{scalar}
\eea
where $\nabla_A$ is a covariant derivative with respect the base $\theta^A=e_{\beta}^A\,dx^{\beta}$. We go assuming that the scalar field has a simple form given by: $\Phi(z)=cz+const$, where $c$ is a constant. For this choice we find that the nonzero components of the tensor (\ref{scalar}) are
\bea
T_{00}^S=-T_{11}^S=-T_{22}^S=T_{33}^S=\frac{e^2}{2},
\eea
and the trace of this quantity is given by
\bea
T=\rho-3p+e^2.
\eea

With these quantities we can write the field equations (\ref{motion2}) as follows
\bea
f_RG_{AB}&=&\kappa^2\left[(\rho+p)u_Au_B-p\eta_{AB}+T_{AB}^S\right]-\frac{1}{2}\left[\kappa^2(\rho-3p+e^2)+f+f_T(\rho+p+e^2)\right]\eta_{AB}\nonumber\\
&+&f_T\left[(\rho+p)u_Au_B+T_{AB}^S\right].
\eea
Using the nonzero components of the Einstein tensor (\ref{einsteintensor}) we obtain the following field equations
\bea
2(3\omega^2-m^2)f_R+f&=&\kappa^2(\rho+3p)+f_T(\rho+p),\\
2\omega^2f_R-f&=&\kappa^2(\rho-p)+f_T(\rho+p),\\
2(m^2-\omega^2)f_R-f&=&\kappa^2(\rho-p+2e^2)+f_T(\rho+p+2e^2).
\eea
Manipulating these equations we can rewrite
\bea
\kappa^2e^2+e^2f_T&=&(m^2-2\omega^2)f_R,\\
\kappa^2p&=&\frac{1}{2}(2\omega^2-m^2)f_R+\frac{1}{2}f,\\
\kappa^2\rho+f_T(\rho+p)&=&\frac{1}{2}(6\omega^2-m^2)f_R-\frac{1}{2}f.
\eea
Assuming that these equations satisfy the conditions $f_R>0$ and $f_T>0$ we have that the set of equations that satisfy a causal solution is
\bea
&&m^2=4\omega^2,\label{causal}\\
&&f_R-\frac{e^2}{2\omega^2}f_T=\frac{\kappa^2e^2}{2\omega^2},\\
&&\kappa^2p=-\left[\kappa^2\rho_+f_T(\rho+p)\right]=-\omega^2f_R+\frac{1}{2}f.
\eea
We remember that the relation (\ref{causal}) imply in a critical radius $r_c\to\infty$ therefore for this combination of perfect fluid and scalar field the equations above describe a class of equations that no display violation of causality.

We can conclude that the $f(R,T)$ gravity, thus as proposed is a generalization of the $f(R)$ gravity and in this paper we did a generalization for the study of the violation of causality in this generalized model. First, we study the gravitational model for the case in that the content of matter is a perfect fluid and we saw that arises naturally a condition that lead us for the violation of causality and still, we get to observe that the radius critical depend of the gravitational theory and on the content of matter, more specifically of the energy density, as observed in \cite{Reboucas1}, and also of the explicit form of $f_T$, i.e., when we consider the $f(R,T)$ generalized theory we note that the addition of the dependence of the trace of the energy-momentum tensor modify the critical radius for the existence of the closed time-like curves in a G\"{o}del-type universe. Investigating the result obtained  for the critical radius and considering $f_T>0$ it can be observed that the contribution coming of the $f(R,T)$ gravity contributes for decrease the critical radius when compared with the obtained for the case of the $f(R)$ gravity. In the last part of the paper we consider that the content of the matter of the universe is composed of a perfect fluid together with a scalar field and we note that the results found for the $f(R)$ gravity is reproduced for the $f(R,T)$ gravity, i.e., we do not have solutions that display the violation of causality.

\begin{acknowledgments}
This work was partially supported by Conselho Nacional de Desenvolvimento Cient\'{\i}fico e Tecnol\'{o}gico (CNPq) and Coordena\c{c}\~ao de Aperfei\c{c}oamento de Pessoal de N\'{\i}vel Superior (CAPES). A. F. S. has been supported by the CNPq project 476166/2013-6.
\end{acknowledgments}

\end{document}